\documentclass[twocolumn,tighten,times]{aastex62}
\usepackage{amssymb}
\usepackage{amsmath}
\usepackage{amsfonts}
\usepackage{float}
\usepackage{relsize}
\usepackage{verbatim}
\usepackage{color}
\usepackage{comment}
\usepackage{graphicx}
\usepackage{bm}

\newcommand{\todoblank}[1]{}

\graphicspath{{figures/}}

\def\be{\begin{equation}}
\def\ee{\end{equation}}
\def\ba{\begin{eqnarray}}
\def\ea{\end{eqnarray}}

\newcommand{\W}{\ensuremath{\mathrm{W}}}

\newcommand{\SB}{{\mathrm{SB}}}
\newcommand{\CW}{{\mathrm{CW}}}

\newcommand{\R}{{\mathrm{R}}}
\newcommand{\total}{{\mathrm{total}}}

\newcommand{\jitterrefprep}{{(M.~T.~Lam in prep.)}}
\newcommand{\CWrefprep}{{(K. Aggarwal et al. in prep)}}

\shorttitle{Optimizing Pulsar Timing Array Observations}
\shortauthors{Lam}

\begin{document}

\title{
Optimizing Pulsar Timing Arrays Observations for Detection and Characterization of Low-Frequency Gravitational Wave Sources
}
\author[0000-0003-0721-651X]{M.\,T.\,Lam}
\affiliation{Department of Physics and Astronomy, West Virginia University, White Hall, Morgantown, WV 26506, USA; michael.lam@mail.wvu.edu}
\affiliation{Center for Gravitational Waves and Cosmology, West Virginia University, Chestnut Ridge Research Building, Morgantown, WV 26505}

\begin{abstract}

Observations of low-frequency gravitational waves will require the highest possible timing precision from an array of the most spin-stable pulsars. We can improve the sensitivity of a pulsar timing array (PTA) to different gravitational-wave sources by observing pulsars with low timing noise over years to decades and distributed across the sky. We discuss observing strategies for a PTA focused on a stochastic gravitational-wave background such as from unresolved supermassive black hole binaries as well as focused on single continuous-wave sources. First we describe the method to calculate a PTA's sensitivity to different gravitational-wave-source classes. We then apply our method to the 45 pulsars presented in the North American Nanohertz Observatory for Gravitational Waves (NANOGrav) 11-year data set. For expected amplitudes of the stochastic background, we find that all pulsars contribute significantly over the timescale of decades; the exception is for very pessimistic values of the stochastic background amplitude. For individual single sources, we find that a number of pulsars contribute to the sensitivity of a given source but that which pulsars contribute are different depending on the source, or versus an all-sky metric. Our results seem robust to the presence of temporally-correlated red noise in pulsar arrival times. It is critical to obtain more robust pulsar-noise parameters as they heavily affect our results. Our results show that it is also imperative to locate and time as many high-precision pulsars as possible, as quickly as possible, to maximize the sensitivity of next-generation PTA detectors.

 % it may be beneficial to observe only the best-timed pulsars for PTA observations over only a few years whereas for many decades all pulsars contribute significantly%

%different values of amplitude, total time.

\end{abstract}


\section{Introduction}

As the detection of low-frequency gravitational waves (GWs) nears \citep{Taylor+2016}, pulsar timing array (PTA) collaborations must begin looking towards the future characterization of the GW sky. As with ground-based detectors, we must begin planning for the next-generation of PTA detectors, one optimized for these observations. The North American Nanohertz Observatory for Gravitational Waves \citep[NANOGrav;][]{McLaughlin2013} collaboration, one of several efforts worldwide \cite[e.g.,][]{IPTADR1,Desvignes+2016,Reardon+2016}, is currently observing over 70 high-precision millisecond pulsars (MSPs) in its PTA detector for the purpose of low-frequency GW detection from both a stochastic background and from single sources. Without a detection, we have placed constraints on the environments of supermassive black hole binary (SMBHB) mergers, cosmic strings, and inflationary era GWs \citep{NG5BWM,NG5CW,NG11GWB}.

Detector sensitivity depends on the GW signal we wish to observe. A stochastic background requires observations of many MSPs \citep{Siemens+2013,vs2016} whereas the sensitivity to single sources such as from a single binary or merger event requires the highest timing precision from a few of the best-timed MSPs \citep{Ellis+2012}. PTA observations require years to decades of a timing baseline to detect nanohertz-regime GWs and therefore large amounts of telescope time are required. In theory with enough observing time we could observe enough pulsars with adequate timing-precision to cover both science targets (stochastic background versus single source) but practical limitations apply. NANOGrav currently observes its pulsars on a monthly cadence except for a handful of the highest-precision pulsars which are observed weekly, with the goal of covering both stochastic background and single continuous wave (CW) source characterization. However, the efficacy of the approaches to maximizing GW sensitivity has been unclear so far.

%Even with a full observing time at a telescope

In \citet{optimalfreq}, we examined pulse arrival-time uncertainty for MSPs as a function of the radio frequencies observed by specific telescopes taking into account a wide variety of effects. The requirements per pulsar vary but large bandwidths covering much of the radio spectrum typically used for high-precision pulsar timing ($\sim$GHz) are needed to obtain the best possible arrival-time estimates. In this work, we will consider the time-allocation optimization for various pulsars in the array to maximize overall GW sensitivity. \citet{KJ+2012} examined this problem first by considering a simplified PTA and providing the methodology to optimize a specific stochastic-background detection statistic given observing constraints from one or several telescopes.

Here we will develop the methodology for time optimization for the goals of detecting and characterizing both the stochastic background and single CW sources. We apply this formalism to the 45 pulsars presented in the NANOGrav 11-year data set \citep[NG11;][]{NG11yr}, providing specific prescriptions for allocating the observing time per pulsar depending on the two science goals. In \S\ref{sec:statistic}, we describe the cross-correlation statistic used as our GW sensitivity metric. We apply our formalism to the NG11 pulsars in \S\ref{sec:application} and describe future directions in \S\ref{sec:future}. 

%We can only suggest guidelines for a unified optimization scheme for PTAs but ultimately the choice depends on project goals. 

%Throughout this work, we will always specifically denote the term ``radio frequency'' in contrast to GW frequency.

\section{The Cross-Correlation Statistic}
\label{sec:statistic}

%Following both \citet{cs2013} and \citet{Siemens+2013}, we construct a metric for determining the GW sensitivity as a function of sky location. 

For the signal-to-noise (S/N) metric that we wish to maximize, we use the cross-correlation statistic derived in \citet{Siemens+2013}. We first break it into its constituent components for an individual pulsar pair (pulsars labeled with subscripts $i,j$), given as
\be
\rho_{ij}=\left(2T_{ij}\chi_{ij}^2\int_{f_{\mathrm{L}}}^{f_{\mathrm{H}}}df\frac{P_g^2(f)}{P_i(f)P_j(f)}\right)^{1/2},
\label{eq:rho}
\ee
where $T_{ij}$ is the overlapping timespan between the two pulsars, $\chi_{ij}$ is the overlap reduction function (e.g., the Hellings-Downs correlation function for an isotropic background) that depends on pulsar angular separations \citep{hd1983,cs2012}, $f$ is the frequency with $f_{\mathrm{L}}$ and $f_{\mathrm{H}}$ the low- and high-frequency spectral cutoffs, respectively, $P_g$ is the ``signal'' GW power spectrum\footnote{We assume one-sided spectra throughout this work.} (``Earth term''), and $P_{i,j}$ are the total pulsar noise spectra that include the effect of GWs at the pulsar (``pulsar term''). Since power is absorbed by parameter fits in the timing models, we assume a low-frequency cutoff $f_{\mathrm{L}}\approx1/T_{ij}$, due to the spin-period and spin-period-derivative quadratic subtraction from the TOAs to good approximation \citep{Siemens+2013}. Once we compute our individual $\rho_{ij}$ values, the average S/N statistic is given simply by
\be
\left<\rho\right>=\left(\sum\rho_{ij}^2\right)^{1/2}.
\label{eq:avgrho}
\ee
We will drop the angular brackets denoting the average going forward for brevity.

%\footnote{With increasing $T_{ij}$, power absorbed by other parameters decreases significantly, and the approximation improves.}.

% we include transmission or filter functions $\T(f)$ in the integral \citep{bnr1984,Vitale+1997,Madison+2013} \todo{it is not clear this form of a transmission function is correct, this is being worked on}. The fit for pulsar spin period and spin period derivative is typically approximated by assuming the low-frequency cutoff $f_L \approx 1/T_{ij}$ \citep{Siemens+2013,NG9yr} but here we include these in the numerical integration. Following \citep{Madison+2013}, we consider a fit for pulse phase, spin period, spin period derivative, sky position, proper motion, and parallax, and construct $\T(f)$ using the Gram-Schmidt orthonormalization process accordingly. 

%\footnote{note this notation differs from \citealt{cs2013} to avoid replication of the variable $\beta$}

\subsection{Stochastic GW Background}

Here we will consider the the form of a stochastic GW background of the power-law form \citep{Jenet+2006}
\be
P_{\SB}(f)=\frac{A_{\SB}^2}{12\pi^2}\left(\frac{f}{1~\mathrm{yr}^{-1}}\right)^{2\alpha}f^{-3}\equiv bf^{-\beta},
\label{eq:PSB}
\ee
where $A_{\SB}$ is the strain amplitude at a frequency of 1~yr$^{-1}$ and $\alpha$ is the spectral index of the characteristic strain, where when cast in terms of $\beta=13/3$ for SMBHBs. Other values exist for primordial GWs or cosmic strings \citep{Jenet+2006}. \citet{Siemens+2013} provide scaling relations for different regimes of the strength of the signal power spectra versus the noise power spectra but we consider the full form of the Eq.~\ref{eq:rho} integral. The pulsar noise spectra are given by three terms,
\be
P_i(f)=P_{\W,i}(f)+P_{\R,i}(f)+P_{\SB}(f),
\label{eq:pulsarspectrum}
\ee
the sum of the uncorrelated-in-time white noise $P_{\W,i}(f)$, correlated-in-time red noise $P_{\R,i}(f)$, and time-correlated but spatially-uncorrelated GW pulsar-term with a power spectrum still given by $P_{\SB}(f)$. For a single white-noise rms $\sigma_i$, the white-noise term is $P_i(f)=2\sigma_i^2\Delta t_i=2\sigma_i^2/c_i$, where $\Delta t_i$ is the time between observations and $c_i$ is the equivalent cadence which we use as our per-pulsar model parameter discussed later. Pulsars with red noise were modeled in the form of a power law, $A_{\R,i}(f/1~\mathrm{yr}^{-1})^{-\gamma_i}$, otherwise we assumed $P_{\R,i}(f)=0$. Practically, the white-noise levels change over time due to changes in observing bandwidths, integration times, etc., in which case more generally we have (dropping the subscript $i$)
\ba
P_{\W,\total}(f)&=&2\sigma_{\total}^2/c_{\total}\nonumber\\&=&2\sigma_{\total}^2\frac{N_{\total}}{T_{\total}}\nonumber\\&=&2\left(\frac{\sum_nN_n\sigma_{n}^2}{\sum_nN_n}\right)\frac{\sum_nN_n}{\sum_nT_n}\nonumber\\&=&2\frac{\sum_n(T_n/c_n)\sigma_{n}^2}{\sum_nT_n}\nonumber\\&=&\frac{\sum_nP_{\W,n}(f)T_n}{\sum_nT_n}\label{eq:generalPw}
\ea
where $n$ denotes different time periods of length $T_n$, cadence $c_n$, and rms $\sigma_n$, and the number of observations is $N_n=T_n/c_n$.

\subsection{Continuous Wave Sources}

A common method for CW analyses is to perform searches over a multi-dimensional likelihood function that includes intrinsic source parameters (e.g., chirp mass, GW frequency, strain/distance) along with parameters describing the orientation of the binary with respect to the Earth \citep{NG5CW,EPTACW}. Frequentist approaches maximize either an effective matched-filter-type spatially incoherent ($\mathcal{F}_p$) or coherent statistic \citep[$\mathcal{F}_e$;][]{Ellis+2012}. In the upcoming NG11 analysis of CW sources \CWrefprep, we use a Bayesian approach to compute the evidence via a Bayes factor for the parameters while accounting for proper the angular-correlation patterns. We assume here that the correlation must be considered for a detection, as sinusoidal-type waveforms can be present in individual pulsar timing data but due to other systematic effects, such as clock errors or errors in the Solar System ephemerides\footnote{In current NANOGrav work, the ephemerides are accounted for using \textsc{BayesEphem}. The errors are expected to be reduced in the future \citep{NG11GWB}.}. Therefore, the {\it conservative} statistic we maximize over will take a different form than previous calculations (e.g., the $\mathcal{F}_p$-statistic).

% though we will compare to the $\mathcal{F}_p$-statistic as the minimum S/N bound.

Using the cross-correlation statistic (Eq.~\ref{eq:rho}), we can again calculate the average S/N but for CWs. Note that we need to modify the equation by replacing $\chi_{ij}$ with $\widetilde{\chi}_{ij}$, the sky-location-dependent overlap reduction function \citep{Anholm+2009}. The signal spectrum for a source with GW frequency $f_{\CW}$ is\footnote{The boxcar function of length $T$ acts as a finite observing span, which is a multiplication in the Fourier domain of a delta function and normalized sinc function; this procedure can be used to calculate other signal spectra as well. Note the factor of 3 in the denominator versus Eq.~\ref{eq:PSB} from the lack of sky averaging \citep{Anholm+2009,tr2013}}
\be
P_{\CW}(f)=\frac{A_{\CW}^2}{4\pi^2}\delta_{T_{ij}}(f-f_{\CW})f^{-3},
\label{eq:PCW}
\ee
where the approximating function
\be
\delta_T(f)\equiv\frac{\sin(\pi fT)}{\pi f}
\ee
tends to $\delta(f-f_{\CW})$ for infinite time and $A_{\CW}$ is the amplitude of the GW. Source frequencies are expected to evolve over time such that the pulsars experience lower frequencies than at the Earth \citep{NG5CW}. Since the signal term (Eq.~\ref{eq:PCW}) will cause $\rho_{\ij}$ to include power mostly from $f_\CW$, we do not include $P_{\CW}(f)$ in the pulsar noise term though we do include the $P_{\SB}(f)$ term.

\section{Application to the NANOGrav 11-year data set}
\label{sec:application}

We applied our formalism to the 45 MSPs discussed in NG11. Each pulsar noise model contained several white-noise components and a red-noise power-law model when significant. For simplicity, we used the noise parameters directly reported in the paper (see their Table 2). While work has been done to study the many contributions to the TOA uncertainties \citep{sod+2014,NG9WN}, there remain components which have not been well quantified, e.g., from radio-frequency interference or polarization mis-calibration \citep{optimalfreq}. We took the conservative approach and used the directly measured values from the full timing of each pulsar.

We assumed that during the entire timespan encompassing NG11 each pulsar was observed monthly per year per telescope except for weekly observations of two pulsars at the Green Bank Telescope (GBT) starting in 2013 (PSRs J1713+0747 and J1909--3744 for 30 minutes each) and five pulsars at the Arecibo Observatory (AO) starting in 2015 (PSRs J0030+0451, J1640+2224, J1713+0747, J2043+1711, and J2317+1439 for one hour each); see NG11 for more details. The goal of these observations was to increase the sensitivity of the PTA to CWs~\citep{NG11yr}. For the high-cadence campaigns, we assumed a time increase of 30 minutes during the non-monthly-observation weeks at GBT or 60 minutes weekly at AO; the total time was 864 hours per year. Recall that the reported white-noise rms of these pulsars will not be affected because we separate out the observational cadence in $P_{\W,i}(f)$.
 
For our optimization analysis, we assumed that the pulsar white-noise rms values were fixed over all epochs, as well as the parameters describing the red-noise power law; future mitigation of timing effects, for example from radio-frequency-dependent pulse-propagation delays, may alter the measured parameters. We note that after the first several years of observing, we switched to wider-bandwidth telescope backends, causing the effective white-noise rms to decrease (see Eq.~\ref{eq:generalPw}), which we ignore here for simplicity.

% we also ignore gaps in observing and also extra observations in this analysis. 

For some pulsars where the time baseline was relatively short, the rms noise was very small, largely because the timing-model fit removes more power at longer timescales, e.g., from red noise, than for longer baselines \citep{bnr1984,Madison+2013}. For example, PSR J2234+0611 was listed with an rms of 30 ns, the lowest of any pulsar yet it was only observed for two years and neither contains the lowest median template-fitting TOA uncertainties~\citep{NG11yr} nor does it have the lowest rms from other white-noise sources\jitterrefprep. Nonetheless, we used all NG11 pulsar noise parameters as reported with the important caveat that future estimates will likely change.

\begin{deluxetable*}{lccccc|cccccc}
\tablecolumns{12}
\tabletypesize{\scriptsize}
\tablecaption{Optimal Pulsar Cadences for a Stochastic Background\label{table:times}}
\tablehead{
\colhead{Pulsar} & \colhead{Telescope} & \colhead{Timespan} & \colhead{$\sigma_i$} & \colhead{$A_{\mathrm{R}}$} & \colhead{$\gamma$} & \multicolumn{6}{c}{Cadence~(hours/year)}\\
\colhead{} & \colhead{} & \colhead{} & \colhead{} & \colhead{} &\colhead{} &\multicolumn{2}{c}{$A_{\mathrm{SB}}=1\!\times\!10^{-15}$}&\multicolumn{2}{c}{$A_{\mathrm{SB}}=6\!\times\!10^{-16}$} & \multicolumn{2}{c}{$A_{\mathrm{SB}}=2\!\times\!10^{-16}$} \\
\colhead{} &\colhead{} & \colhead{(yr)} & \colhead{($\mu$s)} &\colhead{($\mu$s)} &\colhead{} & \colhead{$T_+ = 10$} & \colhead{$T_+ = 20$} & \colhead{$T_+ = 10$} & \colhead{$T_+ = 20$} & \colhead{$T_+ = 10$} & \colhead{$T_+ = 20$} 
}
\startdata
J0023+0923 & AO & 4.4 & 0.308 & $-$ & $-$ & 18.9 & 15.6 & 19.3 & 17.8 & 12.7 & 17.4 \\
 J0030+0451 & AO & 10.9 & 0.241 & 0.025 & 4.0 & 14.0 & 11.6 & 13.3 & 13.4 & 15.2 & 13.1 \\
 J0340+4130 & GBT & 3.8 & 0.454 & $-$ & $-$ & 13.2 & 12.6 & 12.4 & 12.3 & 4.2 & 11.1 \\
 J0613$-$0200 & GBT & 10.8 & 0.199 & 0.212 & 1.2 & 9.7 & 8.9 & 11.2 & 9.4 & 15.9 & 11.5 \\
 J0636+5128 & GBT & 2.0 & 0.611 & $-$ & $-$ & 23.0 & 26.3 & 17.7 & 26.2 & $-$ & 23.7 \\
 J0645+5158 & GBT & 4.5 & 0.180 & $-$ & $-$ & 16.4 & 15.7 & 17.3 & 15.9 & 16.0 & 17.4 \\
 J0740+6620 & GBT & 2.0 & 0.190 & $-$ & $-$ & 17.1 & 15.8 & 17.1 & 15.9 & 10.6 & 17.5 \\
 J0931$-$1902 & GBT & 2.8 & 0.495 & $-$ & $-$ & 14.9 & 14.5 & 13.9 & 14.3 & $-$ & 11.6 \\
 J1012+5307 & GBT & 11.4 & 0.354 & 0.476 & 1.5 & 8.6 & 8.8 & 9.0 & 9.2 & 10.8 & 10.4 \\
 J1024$-$0719 & GBT & 6.2 & 0.324 & $-$ & $-$ & 12.4 & 11.6 & 13.1 & 11.6 & 11.5 & 12.1 \\
 J1125+7819 & GBT & 2.0 & 0.483 & $-$ & $-$ & 13.5 & 14.8 & 9.9 & 14.6 & $-$ & 13.0 \\
 J1453+1902 & AO & 2.4 & 0.757 & $-$ & $-$ & $-$ & 6.7 & $-$ & 5.7 & $-$ & $-$ \\
 J1455$-$3330 & GBT & 11.4 & 0.571 & $-$ & $-$ & 15.8 & 15.6 & 16.5 & 15.5 & 16.3 & 16.2 \\
 J1600$-$3053 & GBT & 8.1 & 0.181 & $-$ & $-$ & 16.2 & 15.2 & 18.3 & 15.1 & 26.0 & 17.7 \\
 J1614$-$2230 & GBT & 7.2 & 0.183 & $-$ & $-$ & 17.8 & 16.5 & 19.6 & 16.7 & 28.2 & 18.9 \\
 J1640+2224 & AO & 11.1 & 0.382 & $-$ & $-$ & 16.7 & 13.5 & 16.4 & 14.6 & 18.0 & 14.5 \\
 J1643$-$1224 & GBT & 11.2 & 0.757 & 1.619 & 1.3 & 11.2 & 11.2 & 11.8 & 11.9 & 10.8 & 12.4 \\
 J1713+0747 & AO/GBT & 10.9 & 0.103 & 0.021 & 1.6 & 14.1/9.3 & 13.6/8.6 & 15.9/10.5 & 14.8/8.7 & 22.3/16.4 & 15.2/10.5 \\
 J1738+0333 & AO & 6.1 & 0.364 & $-$ & $-$ & 26.4 & 24.5 & 27.2 & 24.7 & 29.0 & 26.3 \\
 J1741+1351 & AO & 6.4 & 0.102 & $-$ & $-$ & 20.6 & 20.2 & 22.4 & 19.7 & 29.8 & 21.4 \\
 J1744$-$1134 & GBT & 11.4 & 0.403 & $-$ & $-$ & 21.6 & 21.4 & 23.5 & 21.7 & 29.8 & 23.2 \\
 J1747$-$4036 & GBT & 3.8 & 1.580 & 1.823 & 1.4 & 7.4 & 14.0 & $-$ & 12.0 & $-$ & $-$ \\
 J1832$-$0836 & GBT & 2.8 & 0.184 & $-$ & $-$ & 21.5 & 19.6 & 23.5 & 19.9 & 30.9 & 22.7 \\
 J1853+1303 & AO & 4.5 & 0.205 & $-$ & $-$ & 35.1 & 35.5 & 36.9 & 35.4 & 45.2 & 38.4 \\
 B1855+09 & AO & 11.0 & 0.482 & 0.069 & 3.0 & 29.1 & 31.4 & 28.1 & 29.6 & 26.9 & 30.3 \\
 J1903+0327 & AO & 6.1 & 0.573 & 1.615 & 2.1 & 18.4 & 19.4 & 20.2 & 22.9 & 17.1 & 24.9 \\
 J1909$-$3744 & GBT & 11.2 & 0.070 & 0.042 & 1.7 & 7.3 & 6.7 & 8.1 & 6.8 & 11.5 & 8.2 \\
 J1910+1256 & AO & 6.8 & 0.515 & $-$ & $-$ & 37.7 & 36.5 & 34.3 & 36.0 & 28.7 & 35.3 \\
 J1911+1347 & AO & 2.4 & 0.054 & $-$ & $-$ & 21.2 & 20.9 & 23.6 & 20.7 & 32.6 & 23.1 \\
 J1918$-$0642 & GBT & 11.2 & 0.297 & $-$ & $-$ & 19.4 & 19.3 & 21.3 & 19.3 & 28.7 & 21.4 \\
 J1923+2515 & AO & 4.3 & 0.229 & $-$ & $-$ & 35.0 & 33.7 & 36.3 & 32.9 & 38.3 & 35.4 \\
 B1937+21 & AO/GBT & 11.3 & 0.110 & 0.157 & 2.8 & 8.1/19.7 & 7.5/20.7 & 9.2/21.8 & 7.7/20.5 & 13.8/29.7 & 9.2/23.7 \\
 J1944+0907 & AO & 4.4 & 0.333 & $-$ & $-$ & 35.7 & 35.9 & 38.2 & 37.4 & 35.6 & 38.5 \\
 B1953+29 & AO & 4.4 & 0.394 & $-$ & $-$ & 32.4 & 33.1 & 32.3 & 32.5 & 27.6 & 31.7 \\
 J2010$-$1323 & GBT & 6.2 & 0.260 & $-$ & $-$ & 17.5 & 16.2 & 19.5 & 16.4 & 25.1 & 18.6 \\
 J2017+0603 & AO & 3.8 & 0.091 & $-$ & $-$ & 24.7 & 25.1 & 26.1 & 24.8 & 34.7 & 24.6 \\
 J2033+1734 & AO & 2.3 & 0.500 & $-$ & $-$ & 38.1 & 39.4 & 34.0 & 38.4 & $-$ & 38.1 \\
 J2043+1711 & AO & 4.5 & 0.119 & $-$ & $-$ & 26.0 & 26.0 & 28.3 & 25.1 & 36.0 & 27.2 \\
 J2145$-$0750 & GBT & 11.3 & 0.304 & 0.589 & 1.3 & 6.4 & 5.9 & 7.2 & 6.3 & 9.2 & 7.8 \\
 J2214+3000 & AO & 4.2 & 1.330 & $-$ & $-$ & 9.3 & 17.4 & $-$ & 14.0 & $-$ & $-$ \\
 J2229+2643 & AO & 2.4 & 0.203 & $-$ & $-$ & 20.9 & 18.6 & 20.9 & 18.7 & 15.1 & 18.1 \\
 J2234+0611 & AO & 2.0 & 0.030 & $-$ & $-$ & 8.6 & 7.0 & 8.1 & 7.1 & 11.8 & 8.3 \\
 J2234+0944 & AO & 2.5 & 0.205 & $-$ & $-$ & 26.9 & 23.2 & 26.9 & 23.4 & 25.5 & 24.4 \\
 J2302+4442 & GBT & 3.8 & 0.836 & $-$ & $-$ & 7.8 & 9.3 & 4.5 & 8.3 & $-$ & $-$ \\
 J2317+1439 & AO & 11.0 & 0.287 & $-$ & $-$ & 18.3 & 18.4 & 17.4 & 17.8 & 16.4 & 17.9 \\
 \enddata
\tablenotetext{}{We assumed $\beta=13/3$. Added timespan $T_+$ is in units of years. Dashed entries formally have values but are too small to be reported.}\end{deluxetable*}

\begin{deluxetable*}{lccccc|cccc|c}
\tablecolumns{11}
\tabletypesize{\scriptsize}
\tablecaption{Optimal Pulsar Cadences for Continuous Wave Sources \label{table:sources}}
\tablehead{
\colhead{Pulsar} & \colhead{Telescope} & \colhead{Timespan} & \colhead{$\sigma_i$} & \colhead{$A_{\mathrm{R}}$} & \colhead{$\gamma$} & \multicolumn{5}{c}{Cadence~(hours/year)}\\
\colhead{} &\colhead{} & \colhead{(yr)} & \colhead{($\mu$s)} &\colhead{($\mu$s)} &\colhead{} & \colhead{M84} & \colhead{M104} & \colhead{NGC~3115} & \colhead{NGC~1316} & \colhead{All Sky} 
}
\startdata
J0023+0923 & AO & 4.4 & 0.308 & $-$ & $-$ & $-$ & $-$ & $-$ & 26.6 & $-$ \\
 J0030+0451 & AO & 10.9 & 0.241 & 0.025 & 4.0 & $-$ & $-$ & $-$ & 29.5 & 1.3 \\
 J0340+4130 & GBT & 3.8 & 0.454 & $-$ & $-$ & $-$ & $-$ & $-$ & 2.1 & $-$ \\
 J0613$-$0200 & GBT & 10.8 & 0.199 & 0.212 & 1.2 & $-$ & 1.9 & 5.7 & 8.8 & 1.4 \\
 J0636+5128 & GBT & 2.0 & 0.611 & $-$ & $-$ & $-$ & $-$ & $-$ & $-$ & $-$ \\
 J0645+5158 & GBT & 4.5 & 0.180 & $-$ & $-$ & 55.6 & 14.8 & 74.7 & 76.6 & 20.5 \\
 J0740+6620 & GBT & 2.0 & 0.190 & $-$ & $-$ & 81.7 & $-$ & 120.9 & 56.4 & 30.2 \\
 J0931$-$1902 & GBT & 2.8 & 0.495 & $-$ & $-$ & 1.1 & $-$ & $-$ & 13.3 & $-$ \\
 J1012+5307 & GBT & 11.4 & 0.354 & 0.476 & 1.5 & 1.9 & $-$ & 2.5 & $-$ & $-$ \\
 J1024$-$0719 & GBT & 6.2 & 0.324 & $-$ & $-$ & 35.7 & 37.5 & 59.1 & 15.5 & 2.8 \\
 J1125+7819 & GBT & 2.0 & 0.483 & $-$ & $-$ & $-$ & $-$ & $-$ & $-$ & $-$ \\
 J1453+1902 & AO & 2.4 & 0.757 & $-$ & $-$ & $-$ & $-$ & 60.0 & $-$ & $-$ \\
 J1455$-$3330 & GBT & 11.4 & 0.571 & $-$ & $-$ & $-$ & 5.0 & $-$ & $-$ & $-$ \\
 J1600$-$3053 & GBT & 8.1 & 0.181 & $-$ & $-$ & 40.8 & 57.5 & $-$ & 16.8 & 32.5 \\
 J1614$-$2230 & GBT & 7.2 & 0.183 & $-$ & $-$ & 36.5 & 66.5 & 5.1 & 13.8 & 36.7 \\
 J1640+2224 & AO & 11.1 & 0.382 & $-$ & $-$ & 20.8 & 6.6 & 34.1 & $-$ & $-$ \\
 J1643$-$1224 & GBT & 11.2 & 0.757 & 1.619 & 1.3 & $-$ & $-$ & $-$ & $-$ & $-$ \\
 J1713+0747 & AO/GBT & 10.9 & 0.103 & 0.021 & 1.6 & 69.2/28.6 & 82.4/30.2 & 99.5/35.3 & $-$/9.0 & 30.2/48.1 \\
 J1738+0333 & AO & 6.1 & 0.364 & $-$ & $-$ & 6.1 & 19.6 & 34.0 & $-$ & $-$ \\
 J1741+1351 & AO & 6.4 & 0.102 & $-$ & $-$ & 116.5 & 119.5 & 141.4 & $-$ & 51.3 \\
 J1744$-$1134 & GBT & 11.4 & 0.403 & $-$ & $-$ & $-$ & 6.5 & $-$ & $-$ & 6.8 \\
 J1747$-$4036 & GBT & 3.8 & 1.580 & 1.823 & 1.4 & $-$ & $-$ & $-$ & $-$ & $-$ \\
 J1832$-$0836 & GBT & 2.8 & 0.184 & $-$ & $-$ & 9.7 & 59.3 & $-$ & 38.4 & 60.8 \\
 J1853+1303 & AO & 4.5 & 0.205 & $-$ & $-$ & 49.5 & 43.0 & 39.7 & $-$ & 25.4 \\
 B1855+09 & AO & 11.0 & 0.482 & 0.069 & 3.0 & $-$ & $-$ & $-$ & $-$ & $-$ \\
 J1903+0327 & AO & 6.1 & 0.573 & 1.615 & 2.1 & $-$ & $-$ & $-$ & $-$ & $-$ \\
 J1909$-$3744 & GBT & 11.2 & 0.070 & 0.042 & 1.7 & 16.4 & 26.8 & 12.0 & 41.6 & 32.3 \\
 J1910+1256 & AO & 6.8 & 0.515 & $-$ & $-$ & $-$ & $-$ & $-$ & $-$ & $-$ \\
 J1911+1347 & AO & 2.4 & 0.054 & $-$ & $-$ & 146.2 & 148.8 & 134.6 & 50.6 & 123.4 \\
 J1918$-$0642 & GBT & 11.2 & 0.297 & $-$ & $-$ & $-$ & 4.5 & $-$ & 4.5 & 13.8 \\
 J1923+2515 & AO & 4.3 & 0.229 & $-$ & $-$ & 22.8 & 15.6 & $-$ & $-$ & 13.3 \\
 B1937+21 & AO/GBT & 11.3 & 0.110 & 0.157 & 2.8 & 6.6/7.0 & 3.7/8.5 & $-$/4.5 & 2.9/3.3 & 11.4/8.2 \\
 J1944+0907 & AO & 4.4 & 0.333 & $-$ & $-$ & $-$ & $-$ & $-$ & $-$ & $-$ \\
 B1953+29 & AO & 4.4 & 0.394 & $-$ & $-$ & $-$ & $-$ & $-$ & $-$ & $-$ \\
 J2010$-$1323 & GBT & 6.2 & 0.260 & $-$ & $-$ & $-$ & 1.7 & $-$ & 15.3 & 17.7 \\
 J2017+0603 & AO & 3.8 & 0.091 & $-$ & $-$ & 70.1 & 68.6 & $-$ & 72.3 & 74.9 \\
 J2033+1734 & AO & 2.3 & 0.500 & $-$ & $-$ & $-$ & $-$ & $-$ & $-$ & $-$ \\
 J2043+1711 & AO & 4.5 & 0.119 & $-$ & $-$ & 29.3 & 29.8 & $-$ & 66.8 & 61.1 \\
 J2145$-$0750 & GBT & 11.3 & 0.304 & 0.589 & 1.3 & $-$ & $-$ & $-$ & $-$ & $-$ \\
 J2214+3000 & AO & 4.2 & 1.330 & $-$ & $-$ & $-$ & $-$ & $-$ & $-$ & $-$ \\
 J2229+2643 & AO & 2.4 & 0.203 & $-$ & $-$ & $-$ & $-$ & $-$ & 59.6 & 20.1 \\
 J2234+0611 & AO & 2.0 & 0.030 & $-$ & $-$ & 10.4 & 5.5 & $-$ & 107.1 & 89.5 \\
 J2234+0944 & AO & 2.5 & 0.205 & $-$ & $-$ & $-$ & $-$ & $-$ & 112.1 & 45.0 \\
 J2302+4442 & GBT & 3.8 & 0.836 & $-$ & $-$ & $-$ & $-$ & $-$ & $-$ & $-$ \\
 J2317+1439 & AO & 11.0 & 0.287 & $-$ & $-$ & $-$ & $-$ & $-$ & 19.3 & 4.1 \\
 \enddata
\tablenotetext{}{We assumed $A_{\mathrm{CW}}=1\times10^{-15}$, $f_{\mathrm{CW}}=0.5\mathrm{yr}^{-1}$, $A_{\mathrm{SB}}=1\times10^{-15}$, and $T_+ = 20$ years. Dashed entries formally have values but are too small to be reported.}\end{deluxetable*}

The global-maximum optimization over pulsar cadences was performed using the Nelder-Mead method \citep{NelderMead} implemented in the Python \textsc{SciPy}  package \citep{NelderMeadscipy} via a basin-hopping algorithm \citep{BasinHopping}. To simplify the parameter search, we started with a several-pulsar array and iteratively added pulsars into the array along with the appropriate number of hours per year per source for each telescope, except for PSRs J1713+0747 and B1937+21 where the appropriate time was added to both. We then found the global maximum for that sub-array and used the solution as a starting guess when adding a new pulsar. This algorithm was robust against the order in which pulsars were added. We assumed no new pulsars were to be added to the array and no changes in total observing time from the program at the end of NG11 would be made (both currently untrue). For PSRs J1713+0747 and B1937+21, we assumed for simplicity that they were jitter- and scintillation-noise dominated versus S/N-limited and so the base $\sigma_i$ was the same at both GBT and AO; this is a good approximation for a majority of observations \citep{NG9WN,NG11yr} but should be revisited in the future.

We chose three values for the strain amplitude $A_{\SB}$ of the SMBHB background: $1\times10^{-15}$ (optimistic), $6\times10^{-16}$ (moderate), and $2\times10^{-16}$ (pessimistic). The most pessimistic lower limit for the background is $\approx1\times10^{-16}$ \citep{db2017,Bonetti+2018} though most estimates suggest that $A_{\SB}$ will be at least several times larger  \citep{Ravi+2015}. We also looked at two different values of the total timespan observed after the end of NG11, $T_+=10$ or $20$ years, for calculating $\rho$. In Table~\ref{table:times}, we show our results in terms of the amount of time per pulsar per year allocated given different values of $A_{\SB}$ and $T_+$.

\begin{figure*}[t!]
\centering
    \includegraphics[width=0.85\textwidth]{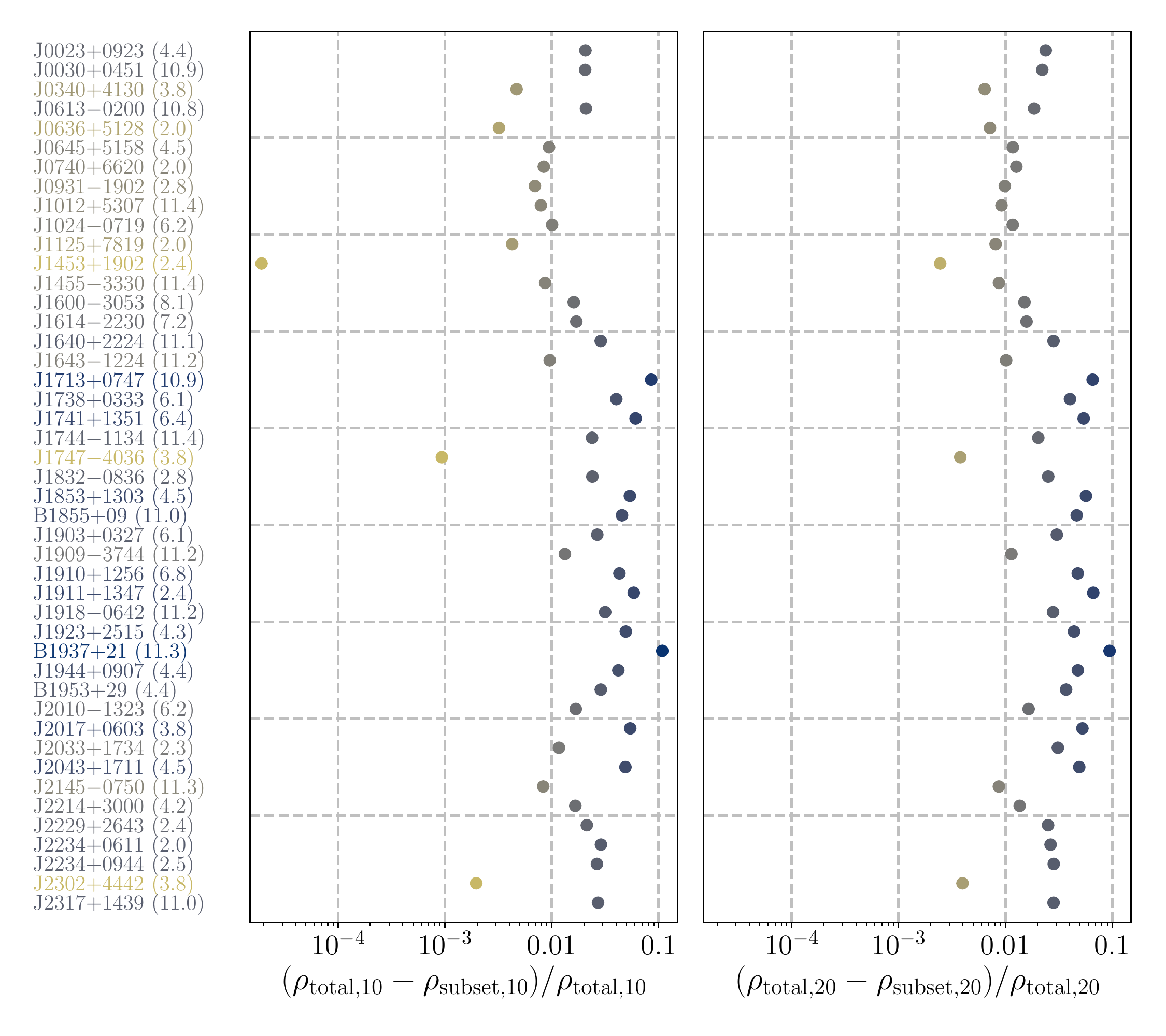}
    \caption{\footnotesize The metric $(\rho_{\mathrm{total}}-\rho_{\mathrm{subset}})/\rho_{\mathrm{total}}$ shows the fractional difference between the S/N of the whole PTA $\rho_{\mathrm{total}}$ and the S/N when the pulsar listed is removed and the time is reallocated and reoptimized $\rho_{\mathrm{subset}}$. The larger the metric, the more important the individual pulsar is in contributing to the PTA's GW sensitivity. Here we assume $A_{\SB}=1\times10^{-15}$. The left panel shows the metric per pulsar when observing for an additional $T_+=10$ years, the right panel for an additional $T_+=20$ years. The total time observed for each pulsar in NG11 is shown in the parentheses (in years). The colors of the pulsar names are the same as of the point in the left panel, which is correlated with the metric.  }
\label{fig:subone}
\end{figure*}

\subsection{Maintaining the PTA: The Time Allocation Per Pulsar}

Table~\ref{table:times} shows that our best pulsars do not overwhelmingly dominate the stochastic-background time allocation. This makes sense when we consider a hypothetical PTA where we ignore all $T_{ij}$ and $\chi_{ij}$ and only have pulsars with white noise. Following \citet{Siemens+2013} but including the differences in white noise and cadences, we have $\rho_{ij}\propto[(c_ic_j)/(\sigma_i^2\sigma_j^2)]^{1/2}$. Then, one can easily show that the total $\rho$ maximizes when the ``quality'' of each pulsar, the white-noise spectrum $2\sigma_i^2/c_i$, is similar, and thus one should spend more time on pulsars with higher white-noise rms. For real PTAs, one should consider red noise, varying time baselines, and sky positions, which will modify the results accordingly though the basic principle still applies. We see that the time allocation is of similar magnitude for all pulsars, though several pulsars do not contribute significantly for certain assumed parameters and thus are not listed (when $c_i<1$~hour/year).

%\todoblank{This is why for example the analysis suggests to add some additional time to PSR~J1713+0747, considered one of our best-timed pulsars \todo{cite}, because the long time baseline will correlate strongly with other pulsars with long baselines.}

%\subsection{Current versus Optimized Stochastic-Background Strategy}

Comparing against a continuation of the NG11 observing strategy versus the optimizations found here, we see a $\sim$5\% increase in $\rho$ for $A_\SB=1\times10^{-15}$ but a nearly $\sim8\!-\!10$\% increase for $A_\SB=2\times10^{-16}$. NANOGrav's general strategy of {\it observing many high-precision pulsars with roughly equal time} (along with high-cadence programs) {\it therefore currently provides close to optimal stochastic-background sensitivity}.

%independent of $T_+$

\begin{figure*}[t!]
\centering
    \includegraphics[width=0.85\textwidth]{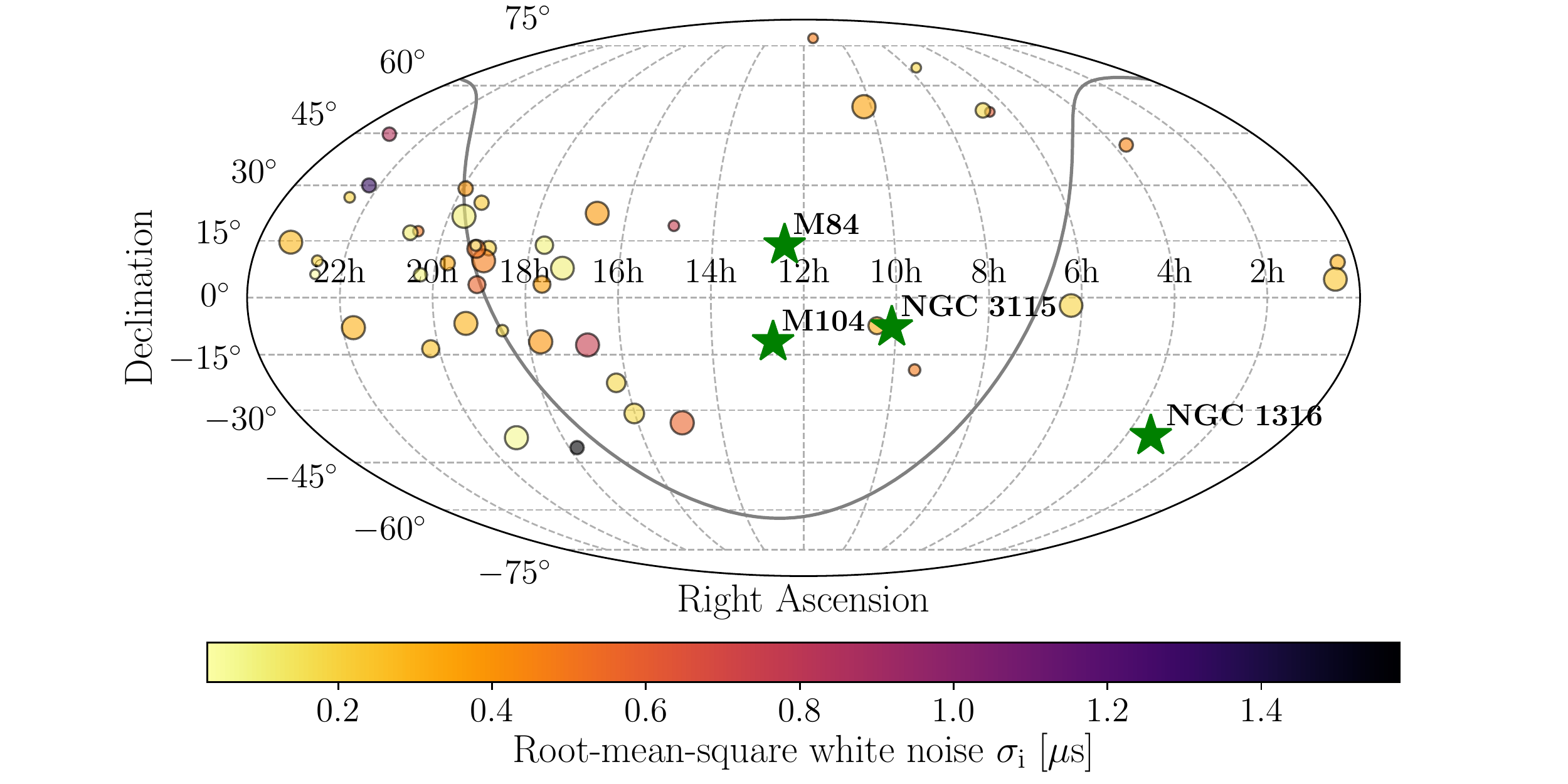}
    \caption{\footnotesize Sky positions of our pulsars (circles) in relation to the four CW sources (stars). Circles sizes are proportional to the timespan observed $T_i$, the colors are proportional to the white-noise rms $\sigma_i$. The curve denotes the Galactic plane. }
\label{fig:skymap}
\end{figure*}

%\todoblank{do a 50+ run?}

\subsection{Trimming the PTA: Removing Pulsars to Improve Sensitivity}
  
Figure~\ref{fig:subone} shows the effect of removing a pulsar from the array and whether reallocating its observing time will increase or decrease sensitivity to the stochastic background. We find that all 45 pulsars contribute significantly for decades-long experiments and that the relative importance of continuing to time these pulsars also tends to improve. However, note in Table~\ref{table:times} where certain pulsars become significant for $T_+=20$~years whereas they are unlisted for earlier times. Formally we find values for these pulsars though they are too small to report given the practical necessity of requiring enough observations to generate a timing solution. The {\it GW significance seems robust to red noise}, as for PSR~J0030+0451 the red-noise index $\gamma=4.0$, which is steeper than the background $\beta=13/3$.
  
%We find that while it may be beneficial to remove pulsars in the short term, e.g., PSR~J1453+1902, all pulsars begin to contribute significantly for longer experiments;  

% \todo{Removing any of the four pulsars written in bold blue improves the GW sensitivity of the PTA when observing for an additional 10 years but not for an additional 20 years; these points are shown by the stars in the right panel for clarity.  }

%\todo{time all pulsars under $\sim$1 us}

\subsection{Directional PTA Tuning}

For CW sources, we looked at specific potential galaxies that may host SMBHBs. \citet{Mingarelli+2017} examined the detectability of individual local SMBHBs by simulating binaries within nearby host galaxies; they estimated that a single source will likely be detectable within $\sim$10 years from now. Assuming different noise statistics of pulsars observed by the International Pulsar Timing Array \citep[IPTA;][]{IPTADR1}, different potential sources may be detectable. Assuming only white noise in the IPTA pulsars, they found that M104 (NGC 4594) may likely host a detectable CW source. In a more updated analysis, they find that M104 is the most likely detectable SMBHB source regardless of the assumed black-hole/galaxy-host scaling relation, followed consistently by M84 (NGC 4374; C. Mingarelli, private communication). NGCs 3115 and 1316 also feature prominently as possible SMBHB-host candidates regardless of pulsar red-noise properties. We looked at these four specific sources independently, shown for reference in Figure~\ref{fig:skymap}, and assumed $T_+=20$~years, a source strain of $A_{\CW}=10^{-15}$ and frequency $f_{\CW}=0.5~\mathrm{yr}^{-1}$, and a stochastic-background pulsar term of $A_{\SB}=10^{-15}$. For comparison, we also considered sources uniformly distributed across the sky with the same assumed parameters and maximized the S/N over all sky positions $\bm{\theta}_n$ as 
\be
\rho_{\mathrm{sky}}=\left(\sum_n\left[\rho(\bm{\theta}_n)\right]^2\right)^{1/2}.
\label{eq:rhosky}
\ee

%  and therefore the observing times for PTA pulsars may be tuned to increase GW sensitivity in that region of the sky

%We do not include a CW pulsar noise term because as the observing time baseline increases we are centered on a single frequency $f_{\CW}$ and the frequency-evolution of the sources move power to slightly different frequencies at the distances of the different pulsars.

%Note that one must consider the sky-location-dependent overlap reduction function, which modifies the usual Hellings-Downs correlation function by removing the assumption of GW isotropy, when computing each $\rho_{ij}$ \citep{Anholm+2009}.

The results of our CW analysis are given in Table~\ref{table:sources}. We find that {\it many pulsars contribute to the sensitivity towards different sources though typically different pulsars are preferred}; this differs from an analysis using the $\mathcal{F}_p$-statistic without sky correlations included in which one or two pulsars dominate the S/N \citep{NG5CW,EPTACW}. Tests using lower $f_\CW$ show a preference for a greater number of pulsars with similar $c_i$, i.e., these sources suggest an observing program more similar to the stochastic background. Recall that the $\rho$ statistic is conservative since it includes correlations but sub-optimal since it does not include signal-waveform matched-filtering. Again, note the caveats above on the uncertainty in the pulsar noise properties, which is why PSRs~J1911+1347 and J2234+0611 contribute significantly in the all-sky analysis. Since M104 and M84 are $\sim$25$^\circ$ separated on the sky, and both $\sim$40$^\circ$ from NGC~3115, we see significant overlap for the pulsars that contribute to $\rho$. For these sources, PSRs~J1744$-$1134 and J1918$-$0642 take up significant portions of observing time in part due to the long timespans and strong correlations with PSR~J1713+0747.

It is possible to ``tune'' PTA observations to efficiently improve the S/N towards multiple targets. In general one can construct a metric that maximizes the sensitivity to a number of possible sources across the sky, potentially in a form similar to Eq.~\ref{eq:rhosky}, and likely should given that any one individual galaxy hosts a detectable CW source is not guaranteed. Current and future PTA experiments may wish to increase their sensitivity to specific sky locations in order to hasten a detection and also improve later characterization, for example in studying the galaxy members of the nearby Virgo cluster in addition to the sources analyzed here.

% Again, note the caveats above on the noise properties of the pulsars, which is why PSR J2234+0611 contributes significantly. Therefore, PSR J2234+0944 contributes as well because of its small separation on the sky; all other pulsars that contribute are observed with the GBT. Since M104 and M84 are $\sim$25 degrees separated on the sky, we see that the same there is significant overlap for the pulsars that contribute to $\rho$. It is possible to tune the array to efficiently improve the S/N towards both targets, and in general one can construct a metric that maximizes the sensitivity to a number of possible sources across the sky, and likely should given that it is not guaranteed that any one individual galaxy hosts a detectable CW source. Current and future PTA experiments may wish to increase their sensitivity to specific parts of the sky in order to hasten a detection and also improve later characterization, for example in studying the galaxy members of the nearby Virgo cluster in addition to these specific sources.

%Since NGC 3115 and M104 are less than 40 degrees separated on the sky, it is possible to tune the array to efficiently improve the S/N towards both targets. 

%; however, other probes of these galaxies may suggest otherwise (for example, see Jones et al. in prep on the non-detection of a candidate in NGC 3115).

\section{Future Applications}
\label{sec:future}

We briefly discuss several next steps for improving GW sensitivity via this framework. One obvious next step is to determine methods of combining the metric provided here that is some weighted combination of different $\rho$ statistics for a stochastic background and specific CW sources, though as various project science goals may differ, we will not expand further.

\subsection{Obtaining Robust Noise Parameters}

As mentioned previously, our results for specific pulsars are skewed because some pulsars with short time baselines have very small rms values. Since the red-noise rms grows with time, it may take many years before a full noise profile is obtained for a given pulsar. Therefore, one can either make a predictive guess of how well a pulsar will perform using various models for red noise \citep[e.g.,][note that there is large scatter in scalings for intrinsic-spin or interstellar-medium-related noise]{NG9EN} or a postdictive recalculation of the different metrics presented here in an attempt to optimize future GW sensitivity.

%\subsection{Directional Tuning}

%As we see from our CW analysis, 

\subsection{Implementation of Wideband Receivers}

While NANOGrav currently observes one pulsar per hour per telescope per epoch, it does so by splitting each observation into two half-hour segments for each frequency receiver. Therefore, 60 minutes of observing time (including overhead) equates to $\lesssim30$ minutes of effective integration time. Wideband receivers will allow for a doubling of the effective integration time for the same on-sky time; in practice a wideband receiver will have a somewhat higher system temperature and an optimization for the best frequency tuning should be performed for maximizing TOA precision \citep{optimalfreq}. While our framework applies generally, with different observational parameters we expect the results of our analyses to differ and one should account for as many inputs as possible in developing observing strategies.

\subsection{International Pulsar Timing Array}

Different PTA collaborations pool their resources into the IPTA collaboration. Each member currently has separate observing strategies. With the resources of many radio telescopes, the best possible GW sensitivity will be obtained by performing the optimization we have laid out across the combined network. Given the differing sensitivities, radio frequency coverages, and cadences of the telescopes, along with practical constraints, it may be preferential for individual telescopes to observe different sources rather than multiple telescopes spending their time to observe the same sources. Future analyses must be performed to investigate how best to optimize all of the IPTA telescopes, especially as new potential member collaborations seek to join.

\acknowledgments

{\it Acknowledgments.} We graciously thank Xavier Siemens, Maura McLaughlin, Stephen Taylor, and Chiara Mingarelli for discussions enabling this work. M.T.L. and the NANOGrav Project receives support from NSF Physics Frontiers Center award number 1430284.

\end{document}